\documentstyle[aps,eqsecnum,epsf]{revtex}

\def\msun{M_{\odot}}

\tighten
\begin{document}
\draft

\title{
Dynamical Evolution of Boson Stars \\
II: Excited States and Self-Interacting Fields}

\author{Jayashree Balakrishna$^1$, Edward Seidel$^{2,3,4}$, and Wai-Mo Suen$^{1,5}$}
\address{${}^1$
McDonnell Center for the Space Sciences \\
Washington University,
St. Louis, MO 63130}

\address{${}^2$
National Center for Supercomputing Applications \\
University of Illinois,
Champaign, IL 61820}

\address{${}^3$
Departments of Physics and Astronomy\\
University of Illinois,
Champaign, IL 61820}

\address{${}^4$
Max-Planck-Institut f$\ddot{u}$r Gravitationsphysik\\
14473 Potsdam, Germany}

\address{${}^5$
Physics Department \\
Chinese University of Hong Kong\\
Hong Kong}
\maketitle

\begin{abstract}

The dynamical evolution of self-gravitating scalar field
configurations in numerical relativity is studied.
The previous analysis on ground state boson stars of non-interacting fields
is extended to excited states and to fields with self couplings.

Self couplings can significantly change the physical
dimensions of boson stars, making them much more astrophysically
interesting
(e.g., having mass of order 0.1 solar mass). The
stable ($S$) and unstable ($U$) branches of equilibrium configurations of boson stars
of self-interacting fields are studied; their behavior under perturbations and their
quasi-normal oscillation frequencies are determined and compared to the
non-interacting case.

Excited states of boson stars with and without self-couplings are studied and
compared. Excited states also have equilibrium configurations 
with $S$ and $U$ branch structures; both branches are intrinsically unstable
under a generic perturbation
but have very
different instability time scales.  We carried out a detailed study
of the instability time scales of these configurations.
It is found that highly excited states spontaneously decay through a cascade of
intermediate states similar to atomic transitions.

\end{abstract}

\section{Introduction}

Various particle physics models suggest that bosons might play
an important role in the evolution of the universe. These models predict
the abundant production of these bosonic particles
in the early universe and their presence in large numbers even today.
Although the
bosonic particles have never actually been directly detected they 
are considered as leading candidates of dark matter[1]. 
These bosons could by a Jeans instability mechanism[2]
condense into compact gravitating objects like boson stars.

Boson stars are made up of self-gravitating complex scalar fields with or
without further self coupling[3,4]. The equilibrium configurations represent
an exact balance between the attractive effect of gravity and the natural
tendency for the scalar field to disperse. The stability of such an object
is hence a central issue. It is well known that equilibrium configurations
of boson stars have stable ($S$) and unstable ($U$) branches as
well as a hierarchy of ground and excited states[3]--[13].
In the first paper in this series[5], the dynamical evolution 
under various perturbations of ground state boson stars
made up of non self-interacting scalar fields was studied.

In the absence of self coupling
the mass
profile of the ground state configurations,
when plotted against the central density
$\phi(0)$, has a peak at $M=0.633m_{Pl}^2/m$ (where $m_{Pl}$
refers to the Planck mass) corresponding to a mass of about $10^{11}$
kg at $\phi(0)_c
=0.271$ (for bosons of mass $m=1$GeV). 
Configurations for which $\phi(0)<\phi(0)_c$ (the $S$ branch) are
stable to perturbations [5,12] while those with $\phi(0)>\phi(0)_c$ (the $U$
branch) are
unstable. Stability here refers to the ability
of a $\phi(0)<\phi(0)_c$ star to settle to a new 
configuration in the same branch under perturbations. A $\phi(0) > \phi(0)_c$ configuration
star is unstable in that, upon perturbations, it cannot stay
on the same branch.  If it cannot lose enough mass and settle to
a stable state it either collapses to a black hole or
disperses to infinity.
Stable boson stars have very specific quasinormal modes
of oscillation under perturbations, a feature important
for the detection and identification of these stars.

In the present  paper the study of paper I[5] is extended to the excited state boson star, and to the
case of a $\lambda \phi^4$ self-interacting scalar field. The dynamical
evolutions of such systems are studied numerically. 

The action
for the system studied in this paper is given by 
 \begin{equation}
 I=\frac{1}{16\pi G}\int d^4 x \sqrt {-g} R - \int d^4 x
 \left[\sqrt{-g} (
 \frac{1}{2}g^{\mu \nu} \partial_\mu \Phi^*\partial_\nu \Phi +
 \frac{1}{2} m^2 \Phi^*\Phi + \frac{1}{4}\lambda |\Phi|^4)\right].
 \end{equation}
There are two reasons to
include the self-coupling interaction[6,8]. Firstly without the $\lambda \phi^4$
term, the maximum mass of a boson star,
\begin{equation}
M=0.633 m_{Pl}^2/m
\end{equation}
(where $m_{Pl}$ refers to the Planck mass and $m$ the mass of the boson)
could be too small to be astrophysically
significant. For example, for $m=$1GeV, $ M = 10^{-19} \msun $ 
where $ \msun $ is the solar mass.
On the other hand, for interacting fields, even with a small coupling,
the mass of the star can be large[6]. In this case
\begin{equation}
M\sim 0.06 \sqrt{\lambda}\frac{m_{Pl}^3}{m^2},
\end{equation}
which is larger than (1.2) by a factor of $\sim \sqrt{\lambda} \frac{m_{Pl}}{m}
\sim 10^{19}\sqrt{\lambda}$, for $m=1$GeV. A moderate value of $\lambda=0.01$
then leads to $M = 0.1 \msun $, which is particularly
interesting due to the
gravitational microlensing data[14].

Secondly boson stars give us a way to study local anisotropy and its effects.
The larger the self coupling parameter $\lambda$ the smaller the fractional anisotropy. Changing
the self coupling parameter for a given central density provides a way to vary
this anisotropy in a natural way[8]. By anisotropy we mean that the radial and tangential components
of the pressure are different. This is of interest because deviations
from perfect fluid assumptions for even nuclear matter is expected
in the presence of strong gravitational fields. In boson stars this
anisotropy appears very naturally. Although we have not 
studied specifically the anisotropies in this paper, it provides motivation for
adding a self-coupling term.

The results of Paper I are also extended by considering
the evolution of excited states with 
and without self coupling. This is of
importance because if boson stars exist and are detected, they are most likely
those interacting with their environment and going through some excitation process.
Excited configurations might also be intermediate stages
during the formation process of these stars. In this study the
configurations considered are spherically symmetric; all perturbations
of equilibrium configurations are purely radial. Full 3D simulations are 
underway and will be reported in a future paper. The remainder of this paper 
is organized as follows:

Section II sets up the mathematical foundations of the problem, including
the equilibrium and evolution equations. The calculation of the initial
data sets are discussed and the techniques used to evolve 
the system numerically are briefly outlined.

Section III details the evolution of ground state
configurations with self coupling. We show that they have similar
$S$ and $U$ branch structures as boson stars without self-coupling.
$S$ branch stars are stable with regard to
perturbations. By this we mean that under small perturbations they
return to configurations on the same branch although not to the
same configurations.
We studied in detail the fundamental quasinormal modes of oscillation
of $S$-branch stars (those for which $\phi(0)<\phi(0)_c$). They are 
important characteristics for observations. They can also be
used to predict the end point of 
evolution of perturbed stars, as well as a comparison between modes for different
self couplings. 
In the
next subsection of section III, the migration of a $U$-branch star (that for
which $\phi(0)>\phi(0)_c)$ to the
$S$-branch is described. The essential features of the
$\lambda=0$ case are retained. If a $U$-branch star is perturbed by the 
addition of mass,
the star will collapse to a black hole.
When, as a result of perturbation, the star's mass is reduced,
corresponding to annihilation of scalar particles, the star
expands and moves to the $S$-branch,
oscillates, and settles
to a new equilibrium configuration of lower mass.

Section IV studies various aspects of the evolution of excited states. 
Excited states have similar band structures to ground state stars.
In this paper, we study generic perturbations that may exist for a 
boson star in an astrophysical environment, e.g., some additional
scalar particles falling in. The S-branch excited states have previously been found to be
stable under infinitesimal perturbations that conserve the total mass and particle number of the boson star[4].
We find that
these stars are inherently unstable irrespective of whether they
lie on the $S$ or the $U$ branch but the time scales of instability are different. This result is consistent with the study of infinitesimal
perturbations[12].
If they cannot lose enough mass to
transit to the ground state they either collapse to black holes or, as in the
case of
stars for which $M>N\,m$ ($M=$ mass of the star, $N=$ number
of bosons and $m=$ mass of one boson) disperse to infinity.
The decay of some higher excited configurations are also studied. These
higher node configurations cascade through intermediate configurations
of lower excited states on their way to collapse. This is reminiscent of
atomic transitions where atoms go from an excited state to lower states
through intermediate ones, lending credence to the idea that boson stars
are like gravitational atoms[3].  A brief conclusion
follows in Section V. An appendix at the end shows some features of the
the high $\Lambda$ ($\Lambda=\lambda/4\pi m^2G$) configurations, including a calculation of quasinormal
modes.

\section{formulation and equilibrium models}
 
In this section the mathematical formulation of the problem, the
creation of the equilibrium and perturbed boson star modes and the
numerical code used to study them are described. The formulation is
the same as that of Paper I in this series[5] except for the
self-interaction term.  The numerical treatment used in this code has
various improvements over that described in Paper I.  Some details of
the numerical code, e.g., convergence tests etc, which are
similar to those reported in Paper I
will not be repeated here.

The action for a self-gravitating scalar field given by equation (1),
leads to the scalar field equation
 \begin{equation}
 g^{\mu\nu}\Phi_{;\mu\nu} - m^2\Phi -\lambda(\Phi^*\Phi)\Phi=0,
 \end{equation}
for the complex scalar field $\Phi=\Phi_1+i\Phi_2$
 and the Einstein field equations
 $$
 R_{\mu\,\nu}-\frac{1}{2}g_{\mu\,\nu}R=8\pi GT_{\mu\,\nu}.
 $$

 The metric for this spherically symmetric system can be written as
 \begin{equation}
 ds^2=-{\bf N^2 dt^2 +\bf g^2 dr^2 + r^2 d\Omega^2},
 \end{equation}
 where $\bf g$, the radial metric, and $\bf N$ the lapse are functions of $(\bf
 t,\bf r)$
 with $\bf r$ being the circumferential radius. 
 This form of the metric is known as the radial gauge. In the absence of a 
 shift vector $\beta ^a$, this form of the metric can be maintained for all
 time by enforcing the
 polar slicing condition. This is a condition on the lapse that requires
 $ K_{\theta\theta}+K_{\phi\phi}=0$ where $K_{ij}$ is 
 the extrinsic curvature tensor. This slicing condition
 causes the lapse ${\bf N}$ to decrease rapidly if an apparent horizon is
 approached[15].

The equilibrium boson star configurations are those in which the metric is time 
independent. The scalar field $\Phi$ itself oscillates with fixed frequency
$\omega_0$:
\begin{equation}
\Phi({\bf t},{\bf r})=\Phi_0({\bf r})e^{-i\omega_0 {\bf t}},
\end{equation}
but due to the $U(1)$ symmetry of the Lagrangian, the stress energy tensor
and the spacetime geometry are time independent.
In dimensionless coordinates we have
\begin{equation}
r= m{\bf r}, \quad t=\omega_0{\bf t},\quad \sigma=\sqrt{4\pi G} \Phi, \quad 
N= {\bf N}\frac{m}{\omega_0},\quad \Lambda=\frac{\lambda}{m^2 4\pi G},
\end{equation}
and the Einstein and Klein-Gordon equations under these conditions are
\begin{equation}
\sigma_0' = \chi_1
\end{equation}
\begin{equation}
\chi_1'= -\left[\frac{1}{r}+\frac{g^2}{r}-rg^2\sigma_0^2\right]\chi_1 -
\left[\frac{1}{N^2}-1\right]\sigma_0g^2+\Lambda ( g^2\sigma_0^3)
\end{equation}
\begin{equation}
g'=\frac{1}{2}\left[\frac{g}{r}-\frac{g^3}{r}+\sigma_0^2rg^3\left[1+\frac{1}{N^2}
\right] + r g \chi_1^2+\frac{1}{2}\Lambda(g^3 r \sigma_0^4)\right]
\end{equation}
\begin{equation}
N' =\frac{1}{2}\left[-\frac{N}{r}+\frac{Ng^2}{r}+\frac{rg^2\sigma_0^2}{N}(1-N^2)
+ rN\chi_1^2-\frac{1}{2}\Lambda g^2 Nr\sigma_0^4\right].
\end{equation}
where $\sigma_0\equiv\Phi_0\sqrt{4\pi G}$. 
A prime denotes $\partial/\partial r$ and an overdot denotes
$\partial/\partial t $. All quantities
in this paper are reported in terms of these 
dimensionless parameters unless explicitly stated otherwise.
Regularity at the origin requires that
$g(r=0)=1$ and
that all other quantities are finite at $r=0$. For the solution to represent
an isolated star, it is required that $\sigma(r=\infty)=0$. 
This constitutes an eigenvalue problem. For each choice of $\sigma(r=0)$, the
above set of equations has a solution only when $ N(r=0)$ takes on certain
values. Different eigenvalues correspond to a different number of nodes in
the solution of $\sigma(r)$. Solutions are also obtained for different
values of the coupling parameter $\lambda$. Different families of 
equilibrium configurations are shown in Fig.\ 1. 
The mass profile of ground state boson stars with self coupling 
has the familiar structure 
seen in non self-interacting fields[5]
which is also found in white dwarfs and neutron stars.
The mass grows to a maximum as
the central density is increased, and then
decreases with further increase in 
central density.
(See Fig.\ 1). 
The maximum mass 
increases with $\Lambda$ but the profile is similar. The expectation
that the branch to the left of the maximum is stable ($S$-branch) while
that to the right ($U$-branch) is unstable, as in the
case of ground state configurations without self coupling, is found to be 
true.
Stability here refers to the ability of $S$-branch stars
to settle to new $S$-branch configurations when
perturbed, a feature of importance for the long term
existence of these stars.
\subsection{Evolution equations}

The configurations described above are time independent, equilibrium 
solutions to the Einstein equations. The aim of this paper is to study
their dynamical properties according
to the coupled Einstein- Klein-Gordon equations. In this paper only spherically symmetric configurations
are studied.
In actual numerical evolution the following set of variables are chosen
\begin{equation}
\psi_1\equiv r\sigma_1,\quad \psi_2\equiv r\sigma_2,\quad \pi_1\equiv
\frac{1}{\alpha}\frac{\partial\psi_1}{\partial t},\quad \pi_2\equiv
\frac{1}{\alpha}\frac{\partial\psi_2}{\partial t},
\end{equation}
where
\begin{equation}
\alpha\equiv \frac{N}{g},
\end{equation}
and the subscripts on $\psi_i$ denote the real and imaginary parts of the
scalar field multiplied by $r$.

In terms of these variables and the dimensionless ones in the previous
section the evolution equations are as follows: The radial metric function
$g$ evolves according to
\begin{equation}
{\dot g}=N(\pi_1\sigma_1'+\pi_2\sigma_2').
\end{equation}
The polar slicing equation, which is integrated on each time slice,
is given by
\begin{equation}
N'=\frac{N}{2}\left[\frac{g^2-1}{r}+r\left[(\sigma_1')^2+(\sigma_2')^2-g^2(\sigma_1
^2+\sigma_2^2)\right]+\frac{\pi_1^2+\pi_2^2}{r}-\frac{g^2\Lambda r}{2}(\sigma_1^2+\sigma_2^2)^2\right].
\end{equation}
The Klein-Gordon equation for the scalar field can be written as
\begin{equation}
{\dot\pi_i}=\alpha'\psi_i'+\alpha\psi_i''-\psi_i\left[gN+\frac{\alpha'}{r}+
\Lambda(\psi_1^2+\psi_2^2)\right],\quad i=1,2,
\end{equation}
\begin{equation}
{\dot\psi}_i=\alpha\pi_i,\quad i=1,2.
\end{equation}
The hamiltonian constraint equation is given by
\begin{equation}
\frac{2g'}{rg^3}+\frac{g^2-1}{r^2g^2}-\frac{\pi_1^2+\pi_2^2}{r^2g^2}-
\frac{\sigma_1'^2+\sigma_2'^2}{g^2}-(\sigma_1^2+\sigma_2^2)-\frac{\Lambda}{2}
 (\sigma_1^2+\sigma_2^2)^2=0.
\end{equation}
It is not solved during the evolution but as it is in principle conserved
by the evolution equations, it is monitored closely as an indicator of the
numerical accuracy of the simulation. For further details see[5].

\subsection{Boundary conditions}

Regularity conditions require that $g(r=0)=1$, and $g,N,\sigma_1$ and
$\sigma_2$ have vanishing first spatial derivatives at $r=0$. To
implement this condition numerically, the range of $r$ is extended to
include negative values; $g$, $N$, $\sigma_1$ and $\sigma_2$ are
required to be symmetric about $r=0$. In addition $\psi_1$, $\psi_2$,
$\pi_1$ and $\pi_2$ are antisymmetric about $r=0$. The antisymmetry
allows the determination of $\phi_i$ at the origin as the first
derivatives of $\psi_i$ at $r=0$.  The value of the lapse function is
fixed at the outer edge on each time slice. Its value at the origin is
determined by integrating Eq.\ (2.15) inward from the outer
boundary. The value of $g$ is determined by the evolution.  The mass
of the star is determined by the value of $g$ at the edge of the grid:
\begin{equation}
M=\frac{1}{2}r\left[1-\frac{1}{g^2(\infty)}\right]\frac{m_{Pl}^2}{m},
\end{equation}
where $m_{Pl}$ is the Planck mass and $m$ is the mass of the boson making up the star.

The boundary condition on the scalar field is an outgoing scalar wave
condition.
However, since the dispersion relation of the massive scalar field is
non trivial,
\begin{equation}
\alpha^2 k^2=\omega^2 - N^2m^2,
\end{equation} 
(where $\alpha=N/g$)
there is no perfect algorithm for the implementation of the outgoing wave condition. 
Here we have adopted a two tier approach:

\begin{itemize}
\item{i)}
 A "sponge" region[5] is constructed by adding a potential term at the outer edge of the
computational domain.
\begin{eqnarray}
{\dot\pi_i}=\alpha'\psi_i'+\alpha\psi_i''- \psi_i\left[gN+\frac{\alpha'}{r}+
\Lambda(\psi_1^2+\psi_2^2)\right]+\frac{V}{N}(\pi_i+\psi_i')&,\quad i=1,2,\\
&r_N-D\le  r \le r_N \nonumber
\end{eqnarray}
where $r_N$ is the $r$ value of the outermost grid point and $D$ is an adjustable parameter
representing the width of the sponge. $D$ is typically chosen to be a few times the wavelength
of the scalar radiation moving out. The extra potential term in the above equation is designed to allow 
waves to propagate outward but damp incoming waves.

\item{ii)}

 At the outermost grid point we require
\begin{equation}
\ddot\psi=-\alpha\dot\psi'- \frac{N^2}{2}\psi.
\end{equation}
This is an exact outgoing wave condition only in the case $m=0$. The
second term on the right hand side represents the finite $m$
correction to leading order (recall $N= {\bf
N}\frac{m}{\omega_0}$). The sponge is designed to absorb the
reflection coming from this approximate outgoing
wave condition. We note that recent work on this problem following a
hyperbolic approach seems to provide a simple and more accurate
outgoing boundary condition[16].
\end{itemize}

\section{Dynamical evolutions of perturbed ground state stars with self-interaction}

\subsection{Nature of perturbations}

We study the dynamical properties of the boson stars 
by perturbing the equilibrium field
distribution. The
accretion or annihilation of scalar particles is simulated by the addition of
field in the outer regions of the star or by decreasing it in denser regions of
the star respectively. Another type of perturbation that has been  effected is 
changing $\dot\psi_1$ and $\dot\psi_2$ of the equilibrium
configuration. This perturbation changes the kinetic energy
density distribution. In either case the changes in the metric functions
$g$ and $N$ are determined by the 
constraint equations
and the polar-slicing condition
(integrating Eqs.\
(2.12) 
and
(2.15) on the initial slice.) 
The magnitude and
the length scale of the perturbations can be chosen arbitrarily. The perturbations
are always spherically symmetric.

In the rest of this section we present results obtained on the dynamical evolutions
of ground state equilibrium configurations perturbed in this manner.
Evolutions of both
$S$-branch and $U$-branch stars with and without self coupling are
considered.
\subsection{$S$-branch perturbations}

As shown in Ref.\ [5], in the free field ($\Lambda=0$)
case, a perturbed $S$-branch star oscillates with a definite frequency,
losing mass through bursts of scalar radiation at each expansion and
finally settles to a new $S$-branch configuration of lower mass. 
Here the effect of a self-interacting $\Lambda$ term
on this behavior is examined. In Fig.\ 2
a typical example of the perturbed field configuration and 
radial metric of a star of $\Lambda = 10$ with a central density
of $\sigma (0)=0.1$ is shown. This star has been perturbed by 
accretion of scalar particles in a region of
lower density. Its evolution is detailed below.

Fig.\ 3a shows the radial metric as a function of distance for the same
configuration at various
times. The labels $ A,B,C,D$ correspond to times (in units of the inverse
of the underlying scalar field frequency)
$t=192,306,391,505$ respectively. The positions
of the peaks are labeled $R_0,R_A,R_B,R_C,R_D$, where $R_0$ is the
position of the initial unperturbed peak. Here $R_0=7.95, R_A=8.55,
R_B=6.6, R_C=8.2 $ and $R_D=6.65$ where the 
length scales are in terms of the inverse mass 
of the boson. The oscillations are shown clearly in
Fig.\ 3b which is a plot of the maximum value
of the radial metric as a function of time. The point where this
function is a maximum corresponds
to the core of the star contracting 
to its minimum size in a cycle. Similarly,
the maximum radial metric starts to decrease as the star expands.
At each expansion the star loses mass through scalar radiation. The
oscillations damp out in time as the star starts settling to the new
configuration. A plot of mass vs. time is shown in Fig.\ 3c. The
amount of scalar radiation decreases in time as the oscillations damp out,
as can be seen from the figure. The slope of the curve steadily decreases
as the star starts settling down to its new lower mass 
configuration. The mass is measured at the inner edge of the sponge. Exact details of the 
curves have some dependence on the sponge parameters but the basic results are the same.

A characteristic of the boson star that
could be important for its observation and identification
is its fundamental oscillation frequency which can be determined from Fig.\ 3b.
We found 
$f=1/(199\,N(\infty)) =4.7\times 10^{-3}$. The oscillation frequencies for
a large number of S-branch stars have been compiled in this way.
Fig.\ 4a shows a plot of the oscillation 
frequency versus mass for many slightly perturbed configurations
(masses within 0.1\% of the unperturbed mass.) As the mass increases, 
the frequency increases and then
drops down as the transition point 
($\frac{dM}{d\sigma(0)}=0$)
is approached,
signaling the onset of 
instability. This is seen for both the non-self interacting as well as the self-interacting case.
These quasinormal modes of oscillation characterize $S$-branch 
stars. The point of transition from the $S$ to the $U$ branch
corresponds to a zero frequency of oscillation[12]. 

For a given mass higher $\Lambda$ $ S $-branch stars have a lower
oscillation frequency than similar mass lower $\Lambda$ stars, unless
one is near the transition point of the lower $\Lambda$
configuration. This is not too surprising, since for a given mass the
radius of the star increases with increasing $\Lambda$. We have seen
this trend even for $\Lambda$ values as high as $1600$ (see discussion
on high $\Lambda$ stars in the Appendix).  However since the maximum
mass of higher $\Lambda$ configurations is greater than that of lower
$\Lambda$ configurations their maximum oscillation frequency could be
greater than that for lower $\Lambda$ stars. (This can be understood
as a size effect: on the $S$-branch, higher mass stars are smaller and
have higher frequencies.)  This can be seen in Fig.\ 4a.  The maximum
oscillation frequencies of $\Lambda=5$ and $\Lambda=10$ configurations
are higher than that of the $\Lambda=0$ case. As the stars get much
larger, though, the highest frequency starts to decrease. For example
the maximum frequency for $\Lambda=30$ stars being less than that of
$\Lambda=15$ stars which is in itself lower than that of the
$\Lambda=10$ case. A perturbation calculation for the high $\Lambda$
case is shown in the appendix to show the dependence of quasinormal
mode frequencies on $\Lambda$ for high $\Lambda$ configurations. The
frequency is proportional to the inverse of the square root of
$\Lambda$. Thus as the stars get really big they oscillate less and
less rapidly and numerically it is no longer feasible to evolve them.
(The time step used in the numerical simulation cannot be increased
as it is determined by the intrinsic oscillation time scale of the
scalar field, which is many orders of magniture shorter than the
oscillation time scale of the whole star for these cases.)

The quasinormal mode curves are also
useful in determining evolutions
of strongly perturbed $S$-branch stars and the final configurations they
could settle into. A
perturbed star loses 
mass and settles to a final configuration corresponding to a position on the solid line in the figure. 
In Fig.\ 4b we single out the $\Lambda=10$ curve and plot the evolution of the $S$-branch star
discussed above. The
points $ P1,P2,P3 $ show the route to a new configuration. These
points correspond to
times $t=0, 1200$ and $4800$ respectively. By extrapolating this line to where it meets
the $\Lambda=10$ curve one could expect a 
final mass of about $.76\,m_{Pl}^2/m$ .

\subsection{$U$-branch perturbations}

For the case of non-self interacting fields it was shown in Ref.\ [5] that 
accretion of scalar fields causes $U$-branch stars to collapse to black
holes. However, lowering the density can make the star migrate to the
$S$-branch. These features
are also seen in the presence of self coupling.
Fig.\ 5 shows a migrating $\Lambda=30$ star whose unperturbed overall
field density $\sigma$
has been decreased by about 10\%. Fig.\ 5a shows the behavior
of the radial metric in time as a function of radius. It oscillates about the final $S$-branch configuration
that it will settle into. This final state is shown on the plot as a dark line.
Fig.\ 5b shows the maximum radial metric as a function of time.
The star initially expands
rapidly as it moves to the $S$-branch. This can be seen from the sharp drop
in the radial metric. Once it moves to the stable branch, it oscillates about 
the new configuration that it is going to settle to. Fig.\ 5c shows the mass of the star as a
function of time. It loses mass at each expansion losing less and less mass at each subsequent
expansion and the curve gets smoother and smoother as it prepares to settle to its final state.

Fig.\ 6, which shows the oscillation frequency as a function of mass for $\Lambda=30$, can be
used to predict the end point of migration.
Points $ Q1,Q2,Q3$ and $Q4$ show the migration of this star. These correspond to times of 500,
1000, 2000  and 3500. The oscillation is clearly damping out. The final
configuration it is expected to settle down to is shown as a dot and
corresponds to a stable star of central density 
$\sigma(0)=0.0817$ with a mass of $1.037\, m_{Pl}^2/m$. This example is
typical of a number of simulations of $U$-branch ground state
configurations with self coupling. 

For higher central density stars on the $U$-branch, the 
mass versus central density curve has a second, gentler peak, similar to the 
white dwarf neutron star situation. One might suspect that this corresponds
to another stable and unstable branch respectively.
However we find that configurations on both sides of the peak are
unstable. These configurations always disperse upon perturbation,
consistent with the fact that they have
$M\,>\,N\,m$, where $M$ was the mass of the star and $N\,m$ was the
number of bosons multiplied by the mass of a boson.

In summary ground state configurations of boson stars with self coupling have
stable ($S$) and unstable ($U$) branches just like boson stars without self
coupling. 
The stable configurations have very specific quasinormal modes of
oscillation. The addition of a self-coupling term serves to increase their mass.
The cases considered so far correspond to $\Lambda\sim10^2$ or less.
Appendix $A$ deals
with very high $\Lambda$ stars which need a different technique
due to the very different timescales involved in these
evolutions.

\section{Evolutions of Excited States}

Excited states of boson stars have field configurations characterized by nodes. The first excited state has
one node, the second has two and so on. 
Studying their stability and the time scale
of decay is important in determining the likelihood of finding them in nature.

The mass profiles of excited state boson stars are 
similar to ground state stars. 
Fig.\ 7 shows the mass versus central density curves for ground, first and second
excited states of boson stars without self coupling. The maximum mass increases with the number of
nodes as expected. The similarity of mass profiles of 
excited boson stars to their ground state counterparts, might lead one to expect stable and unstable configurations to the
left and right of the maximum mass respectively in analogy with ground state configurations.
However, our numerical studies show that the
excited boson star configurations on both sides of
the peak are inherently unstable except that the time scales for
instability are different.
If they cannot lose enough mass
to go to the ground state they become black holes or totally disperse. This occurs even if no
explicit perturbations are put in the numerical evolution other
than those introduced by the finite differencing error in the
numerical integration. We have also carried out perturbations that correspond
to more scalar particles falling on to the star, or those that
decrease the scalar field strength
at the center point corresponding to scalar particles decaying through
some channels[5]. The instability shows up in all cases
studied. We note that this instability is $NOT$ in contradiction with the
result of[4], which concluded that $S$ branch excited states are stable under 
infinitesimal perturbations that strictly conserve $M$ and $N$,
(where $M$ is the
mass of the star, $N$ the number of bosons and $m$ the boson mass.) 
Our result of instability under generic perturbation is consistent with the
studies of $\Lambda=0$
stars under infinitesimal perturbation[12]. 
The presence of a self coupling term
increases the time scale of instability
but the essential pattern remains the same. 
In the following we will first give a detailed account of the 
dynamical evolution and
carry out a study of the instability time
scales.

Excited state stars with masses less than the maximum mass of a {\it ground}
state configuration always form ground state stable configurations. In fact
even stars with masses somewhat greater than this can lose enough mass during
their evolution so as to go to the ground state. 
Fig.\
8a shows a perturbed 1-node star whose
mass has been reduced by about $8\%$ to
$0.9\,m_{Pl}^2/m$ by a perturbation
making a transition to the ground
state although the mass is greater than the maximum ground state mass
of $0.633 m_{Pl}^2/m$.  A substantial
amount of scalar radiation is
emitted in the dynamical evolution, which brings 
the mass below the critical value.
The evolution of the radial metric function is
shown.  Although the plot is shown only to a radius $r=100$,
the actual evolution was carried out to $r=300$. The two peaks at $t=0$
is indicative of a first excited state. One of the peaks disappears
gradually as the star goes to the ground state. The star then
oscillates about the ground state configuration that it will finally
settle into.  In Fig.\
8b
we show a 3 node configuration with a total mass
of $0.92 m_{Pl}^2/m$,
going to the ground state
after radiation by scalar waves carries off the excess mass and kinetic
energy.
We have plotted the density function against the radius and time of
evolution.
By the density function we mean density $\rho$ multiplied by an $r^2$
factor which is the mass per $dr$ at radius $r$. $\rho\,r^2$ has $n+1$
maxima for an $n$ node star and hence here we have 4 sets of lines initially.
At the end of the simulation, we see that it settles down to a ground
state configuration with small oscillations with ever decreasing
amplitude.  For these simulations of low central density stars
we put in an explicit perturbation to the equilibrium configuration
since the instability time scales are extremely long without that.

For stars with higher central density, there is 
a critical density above which the stars cannot lose enough mass to go to
the ground state but collapse to black holes. 
In our numerical simulation for one node stars this critical density is
$\sigma(0)=\sigma_2=0.048$. As the central density is increased the
kinetic energy of the highly compressed initial equilibrium configuration
is increased and the star first expands before the eventual collapse
to a black hole.  As the central density  further increases towards the 
$M=N\,m$ point 
the expansion phase becomes longer.
In Fig.\ 
9a
we show the density function ($\rho \times r^2$) against radius
at various times for 4 configurations (an $S$ branch configuration with
$\sigma(0)=0.1$ and three $U$ branch configurations of central densities
$\sigma(0) = 0.3$, $\sigma(0)=0.4$ and $\sigma(0)=0.5$). 
The initial configurations are the equilibrium ones without any
explicit perturbation (except those introduced by the discretization
used in the numerical simulations).
The first frame shows
the $S$ branch star radiating a  little as it makes a
transition to the ground state.
However it cannot sustain this state for long and it collapses to
a black hole.
The time of decay decreases in the case of a $U$ branch stars of
central density $0.3$ and $0.4$. However for the $\sigma(0)=0.5$ star the
star is clearly more dispersive than the previous ones. It
goes through an expansion phase initially although
it finally collapses to a black hole.
Configurations with $\sigma(0) > 0.54$ have
$M>N\,m$. They do not collapse to black holes but disperse to
infinity.


Next we turn to 
a study of instability time scales.
In simulations where a black hole will form, the imminent development of 
an apparent horizon leads to a rapid collapse 
of the lapse due to the polar slicing
used in the evolution.
We take the time of collapse of the lapse at
the origin
to $\sim 10^{-6}$ of its initial central value
to be the
approximate time for formation of the black hole.
Fig.\ 
9b shows this time scale for a 1-node
star without self coupling. 
We plot the decay time scales of first excited state stars as a 
function of central field density.  Again no explicit perturbation is
applied in these evolutions other than the discretization error
in the simulation.
In order to make a fair comparison of the time scale due to
such a perturbation we cover the
radius of the star in all cases by the same number of grid
points. 
The maximum ground state
mass for stars without self coupling is around .633\,$m_{Pl}^2/m$. This
corresponds to a central density of $\sigma_1=0.021$ for a 1-node star. 
We described earlier that
stars with central densities below $\sigma_2=0.048$ (mass of $0.91
m_{Pl}^2/m$) lose enough mass and move to the ground state. Beyond
that and upto the $M=N\,m$ point they collapse to black holes,
while stars with $M > Nm$ corresponding to a central field
density of $\sigma(0)>0.541$ disperse to infinity. The time by which
collapse takes place to a black hole 
decreases with increasing central density along
the $S$ branch. This trend continues for a while into the $U$ branch (starting
at $\sigma=0.25$ until $\sigma\sim 0.4$)
but as one approaches the $M=N\,m$ point
the stars 
lose a significant amount of  matter to infinity
before they collapse to black holes and evolve on a longer time scale.
For example
a star of central density $0.5$ has an initial
radius of $r\sim9$. 
(The radius of the star is defined as the radius which
contains $95\%$ of the mass of the star.)
Its radius increases to as much as 115, more than an order of magnitude,
before it starts collapsing. Dispersion time scales
of a couple of stars for $M>Nm$ which disperse to
infinity (instead of collapsing to
a black hole) are also shown on the figure. To give
a sense of the instability time scales of these
configurations we take the time scale to be
the time by
which these stars disperse to ten times their original radii.
This time drops drastically for stars with $M>>Nm$. 

Next we turn to the case of $\Lambda \ne 0$.
Fig.\ 
10a shows black hole formation for a $\Lambda=30$ star in the first
excited state, with a central density $\sigma(0)= 0.1$. This star had an initial
radius of about 20.7 where the radius is again
defined as that containing $95\%$ of the
mass. The lapse finally collapses to zero indicating that an 
apparent horizon is about to form. 
The time scale of collapse
to the black hole was around 1985 compared to a
time scale of less than 800 for a $ \sigma(0)=0.122 $ star of similar radius
without self coupling which is shown in Fig.\ 
10b.
This is to be expected as the
$\Lambda$ term represents a repulsive force.
The time
is again determined by the lapse collapsing to $10^{-6}$ of its
original value at $r=0$.

We now turn to the evolution of a highly excited state. In Fig.\ 
11a we show the initial field
configuration of a star containing 5 nodes.
For a five node star the density has a central maximum, and then five local maxima, each subsequent one smaller than the one
preceding it. This star is then evolved without any perturbation
except those introduced by the discretization
error of the numerical evolution. 
In Fig.\ 
11b we show a
contour plot of the evolution of the density function in time.
$\rho\,r^2$ has $n+1$
maxima for an $n$ node star and hence here we initially
have 6 sets of lines
centered at dimensionless $r=5, 13, 35, 52$ and $75$ respectively. 
This star has central density $\sigma(0)=0.075$ and it collapses to
a black hole after a long evolution. In the process we see intermediate
states with fewer numbers of nodes.
For comparison in Fig.\ 
11c
show the contour plot 
for the 5 node
star 
upto a time of $t=1000$ 
at which time it has decayed into a four node
state, against the equilibrium density function
of a 4 node star with central density $\sigma = 0.06$.
Very clearly the maxima are at similar radii (although the size
of the peaks are somewhat different).

This feature of
nodes disappearing and the star cascading through lower excited states
is characteristic of decay of
higher excited states of boson stars. Although
in this case and at this time the decaying star is close
to a specific lower excited state, in general the
decaying star is roughly a combination of
lower excited configurations. This is similar
to the decay of atoms in excited states.
However, we note that for the "gravitational atom"[3] there
is no exact superposition
due to the intrinsic non-linearity of the
system.

\section{Conclusion}

In the
first paper in this series, the behavior of boson star ground state
configurations under various perturbations were reported. In this paper the
study has been extended to include boson stars of self-interacting field
and
also
the behavior of boson stars in the excited states. 

The self-coupling term
is important as it can have dramatic effects on the
the mass of the boson stars[6] leading to boson stars of the order of a
solar mass. The mass
profile retains the features of boson stars without self-coupling, having
a central maximum with a stable branch and an unstable branch. All
configurations to the left of the central maximum in the mass vs. central
density curve (see Fig.\ 1) are stable. Under 
small perturbations they have very specific quasinormal modes of oscillation
as seen in Figs.\ 4 and 6 and under perturbations they settle down to new configurations on the same branch. Configurations
that lie on the unstable or $U$-branch, either migrate to new configurations
on the $S$-branch or collapse to black holes, when perturbed. These are
characteristics shared by boson stars with or without
self-coupling .  

Excited states are configurations with nodes.
The field of an $ n^{\rm{th}} $
excited state star has $n$ nodes and its radial metric has $n+1$ peaks.
Their mass profiles are similar to the profiles of boson stars in the
ground state, which makes it appear as if they have a stable and an unstable
branch of configurations. However, irrespective of which branch they lie on,
excited boson stars are unstable with different
instability time scales. Low density excited stars 
having masses close to ground state configurations
will form ground state boson stars after evolution. Denser configurations
form black holes with the decay time decreasing with increasing central
density till one approaches the density corresponding to zero 
binding energy. As the central density approaches this central
density the kinetic energy of the star starts to increase as it
becomes more dispersive. It still collapses to a black hole
but on a larger time scale. Beyond this point for densities corresponding
to positive binding energy the stars disperse to infinity.
We studied the time scales of their instability in figure
9.

An interesting feature in the collapse of excited state boson stars is that
they cascade through intermediate states, during this process,
rather like atoms transiting from excited states to the ground state,
suggesting that boson stars behave in some ways like gravitational atoms[3].
However, an investigation of the possible decay channels (selection
rules) seems much more difficult (if at all possible or meaningful)
here, due to the intrinsic nonlinearity of the theory.
In this paper we have reported evolutions of spherically
symmetric configurations. We are currently extending the study to full $3D$
without spherical symmetry. The numerical study of $3D$ boson stars
in addition to being an
interesting physical problem is also a testbed for 3-dimensional numerical
codes, that enable us to study compact self-gravitating objects without having to deal with hydrodynamic sources as in neutron stars, and singularities as in the
case of black holes.
In particular we aim to study the general two body
problem in relativity by evolving two 3-dimensional scalar field
configurations. The inspiral coalescence of such systems could have 
interesting physical implications as the gravitational wave emitted does
not sensitively depend on the internal structure of the compact objects
until the late stages of coalescence.
Studying the 1-d behavior has been an important tool in testing our 3-d codes,
providing testbeds in the spherically symmetric limit.

\section*{acknowledgements}
We are happy to acknowledge helpful discussions with Greg Daues
and Malcolm Tobias. 
This research is supported in part by the McDonnell 
Center for Space Sciences, the Institute of Mathematical Science of
the
Chinese University of Hong Kong,
the NSF Supercomputing
Meta Center (MCA935025) and National Science Foundation (Phy 96-00507).

\appendix
\section{High $\Lambda$ Case-- Equilibrium Configuration and Quasinormal Mode Determination}

While calculating the eigenvalue for the equilibrium boson star it is found that it gets 
increasingly difficult to calculate the eigenvalue as the value of $\Lambda$ gets large, because
the equations get very stiff. There are two scales to the
problem: a scale of slow variation of the field inside a certain radius
related to $\Lambda$ followed by rapid 
decay outside it. This makes an effective surface layer to the star making
it more similar to neutron stars.
It turns out that the large $\Lambda$ limit can be treated using a
set of approximate equations that are exact in the $\Lambda=\infty$ limit[6]. By making the
following change of variables:
\begin{equation}
{\bar r} =  r/\sqrt{\Lambda}, \quad \quad {\bar \sigma} = \sqrt{\Lambda}\,\sigma
\end{equation}
the equilibrium equations reduce to 
\begin{equation}
\frac{1}{\Lambda}{\bar \sigma}''= -\frac{1}{\Lambda}\left[\frac{1}{{\bar r}}+\frac{g^2}{{\bar r}}-
{\bar r}g^2{\bar \sigma_0}^2\right]{\bar\sigma'} -
\frac{1}{\sqrt{\Lambda}}\left[\frac{1}{N^2}-1\right]{\bar \sigma_0}g^2+
\frac{1}{\sqrt{\Lambda}} ( g^2{\bar\sigma_0^3})
\end{equation}
\begin{equation}
g'=\frac{1}{2}\left[\frac{g}{{\bar r}}-\frac{g^3}{{\bar r}}
+{\bar \sigma_0}^2{\bar r}g^3\left[1+\frac{1}{N^2}
\right] + \frac{1}{\sqrt {\Lambda}}{\bar r} g {{\bar \sigma_0'^2} }
+\frac{1}{2}(g^3 {\bar r} {\bar \sigma_0^4})\right]
\end{equation}
\begin{equation}
N' =\frac{1}{2}\left[
-\frac{N}{{\bar r}}
+\frac{Ng^2}{{\bar r}}
+\frac{{\bar r}g^2{\bar \sigma_0}^2}{N}
(1-N^2)
+\frac{1}{\sqrt{\Lambda}} {\bar r}N{\bar \sigma_0'^2}-\frac{1}{2} g^2 N{\bar r}{\bar \sigma_0}^4\right].
\end{equation}
where the primes refer to differentiation with respect to ${\bar r}$. In the limit of $\Lambda =\infty$,
one can keep terms to leading order in $\frac{1}{\sqrt \Lambda}$ in (A2)--(A4).
In particular (A2) reduces to
\begin{equation}
N=({\bar \sigma}^2 + 1)^{-\frac{1}{2}}.
\end{equation}
However, we note that this is valid only
for a ground state configuration. For a state with nodes it would not be
reasonable to neglect derivative terms compared to terms proportional
to ${\bar \sigma}$, which is zero at a node.

To get an estimate of the accuracy of the high $\Lambda$ approximation, we
compare the solution using the approximate equation for high $\Lambda$ to the
brute force numerical solution of the complete set of equations, for a
$\Lambda =800$, $\sigma=0.05$
boson star in the ground state in
Fig.\ 12. 
The agreement of the fields is quite good
until the outer region where the approximate equations cause the field to abruptly fall to zero.
Comparisons of the radial metric and the lapse are also shown.
In the $\Lambda=\infty$ limit the approximate equations are exact and the 
star really has an outer surface reminiscent of a neutron star. In fact, an equation of state can be
written[6]. In Fig.\ 13 the mass and particle number 
versus central density $\bar {\sigma}$ for high $\Lambda$ stars is shown. One expects as in
the case of other ground state configurations that the configurations with
$M > N_pm$ disperse when perturbed while those on the $U$ branch with $M < N_pm$ would be unstable and
if unable to migrate to the $S$ branch under perturbations would form black holes. (Here we use the symbol $N_p$ for the particle number and not $N$ so
as not to confuse it with the lapse as both these functions figure prominently
in the analysis that follows.)
The particle number is calculated from the current $J^0$ and is given by
\begin{equation}
N_p = 4\pi\int\,r^2\frac{g}{N}\,\omega\, \sigma ^2dr.
\end{equation}
Here $\dot \sigma$ has been replaced by $\omega\,\sigma$ and $d^3 r$ by $4\pi r^2dr$ for the spherically
symmetric case. In terms of the $bar$ coordinates we see that ${\bar N_p} = \sqrt{\Lambda} N_p$.

Next we turn to the determination of the quasi-normal frequency (QNM) of the
high $\Lambda$ stars. In principle one could determine 
the QNM using the dynamical studies as performed 
for the $\Lambda=0$ case. However the procedure is
extremely computationally expensive. 
The scalar field has an inherent oscillation of about $2\pi$ and 
the evolution time steps must be small enough to resolve it. However the code must run long enough
to see a few metric oscillations in order to determine the quasinormal mode. As 
$\Lambda$ gets large, the sizes of the stars also get
large leading to a lower frequency of oscillation. In order to determine the QNM we use instead the
following perturbation analysis based on[8,11] but using our notation
for lapse, fields, time, radius and self-coupling as defined in section II.

We write the perturbed fields as:
\begin{equation}
\sigma =  (\sigma_1 + \iota \sigma_2)e^{\iota \omega t},  \qquad g=g_0+\delta g, \qquad N=N_0+ \delta N
\end{equation}
where
\begin{equation}
\sigma_1=\sigma_0(r)(1+\delta\sigma_1(r,t)),\qquad \sigma_2=\sigma_0(r)\delta\sigma_2(r,t).
\end{equation}
the Klein-Gordon equation can be written as
\begin{eqnarray}
\sigma_1'' +&(\frac{2}{r} + \frac{N'}{N} -\frac{g'}{g})\sigma_1' +g^2(\frac{1}{N^2} 
- 1-\frac{1}{2}\Lambda{\sigma_1}^2)\sigma_1
-\frac{g^2}{N^2}\ddot\sigma_1 +\\  \nonumber
&\frac{g^2}{N^2}(
\frac{\dot N}{N} 
-\frac{\dot g}{g})\left(\dot \sigma_1+ \sigma_2\right)
-2\frac{g^2}{N^2}\dot\sigma_2.
\end{eqnarray}
The field $\sigma_0$ satisfies the equilibrium equation. 
\begin{equation}
\sigma_0'' +\left(\frac{2}{r} + \frac{N_0'}{N} -\frac{g_0'}{g}\right)\sigma_0' +{g_0}^2
\left(\frac{1}{2{N_0}^2} 
- 1-\frac{\Lambda}{2}{\sigma_1}^2\right)\sigma_0
-\frac{{g_0}^2}{{N_0}^2}\sigma_0 = 0.
\end{equation}
Expanding to first order perturbations using (A9) and (A10) we get

\begin{eqnarray}
\delta{\sigma_1}''&+&\left(\frac{2}{r}+\frac{N_0'}{N} -\frac{g_0'}{g} + 2\frac{{\sigma_0}'}{\sigma_0}\right)
\delta{\sigma_1}'
- \frac{{\sigma_0}'}{\sigma_0}
\left(\frac{g_0{\delta g}'-g_0'\delta g}{{g_0}^2}
-\frac{N_0{\delta N}'-N_0'\delta N}{{N_0}^2}\right) \\ \nonumber
&-& \frac{{g_0}^2}{{N_0}^2}\left(2\delta \dot \sigma_2 +\delta
\ddot \sigma_1\right) +  \frac{2}{{N_0}^2}\left(g_0\delta g - \frac{\delta N}{{N_0}}\right)\\ \nonumber
&-& 2g_0\left(1 + \frac{1}{2}\Lambda {\sigma_0}^2\right)\delta g -{g_0}^2\Lambda{\sigma_0}^2\delta \sigma_1=0.
\end{eqnarray}
From 
\begin{equation}
{\bf R_{\mu\nu}} -\frac{1}{2}g_{\mu\nu}{\bf R} = -8\pi G {\bf T_{\mu\nu}}
\end{equation}
and
\begin{equation}
T_{\mu\nu} = \sigma_{,\mu}^{\ast} \sigma_{,\nu} + c.c. -g_{\mu\nu}\left[g^{\alpha\beta}\sigma_{,\alpha}^{\ast}
\sigma_{,\beta} -(1+\frac{1}{4}\Lambda {\Vert\sigma\Vert}^2){\Vert \sigma\Vert}^2\right]
\end{equation}
(where  $T_{\mu\nu} = 4\pi G {\bf T_{\mu\nu}}$) we get
we get the equations
\begin{equation}
\delta T^{0}_{0}=\frac{1}{2r^2}\left(\frac{2r\delta g}{{g_0}^3}\right)',
\end{equation}

\begin{equation}
\delta T^{1}_{1}=
\frac{\delta g}{{g_0}^3}\left(\frac{1}{r^2}-2\frac{{N}_{0}''}{N_0}\right)-\frac{{\delta N}'}{r{g_0}^2N_0}
-\frac{{N_0}'\delta N}{r{g_0}^2N^2},
\end{equation}
and
\begin{eqnarray}
\delta T^{2}_{2}&=&\frac{1}{2}\left[\frac{2\delta g}{g_0}
\left\{\left(\frac{{N_0}'}{{N_0}}\right)^2 
\right.
+\frac{N_0{N_0}''-{{N_0}'}^2}{{N_0}^2}-
\frac{{N_0}'{g_0}'}{N_0g_0}
+\frac{1}{r}\left(\frac{{N_0}'}{N_0}
-\frac{{g_0}'}{g_0}\right)
\right\}
+\frac{1}{{N_0}^2g_0} \delta \ddot g 
\\ \nonumber
&-& \frac{1}{{g_0}^2}\left(\frac{{\delta N}''}{N_0}\right.
- 2\frac{{N_0}'}{{N_0}^2}{\delta N}'
+\frac{{N_0}''}{{N_0}^2}\delta N - \frac{{N_0}'}{N_0g_0}
\left(\delta g' - \frac{{g_0}'}{g_0}\delta g\right)
- \frac{{g_0}'}{N_0g_0}\left({\delta N}'-
\frac{{N_0}'}{N_0}\delta N\right)  
\\ \nonumber
&+&\left. \left.
2\frac{{N_0}'}{{N_0}^2}\left({\delta N}' - \frac{{N_0}'}{N_0}\delta N\right)
+\frac{1}{r}\left\{\frac{{\delta N}'}{N_0}-\frac{{N_0}'}{{N_0}^2}\delta N 
-\frac{{\delta g}'}{g_0}+\frac{{g_0}'}{{g_0}^2}\delta g \right\}
\right)
\right],
\end{eqnarray}
and the equations
\begin{eqnarray}
\delta T^{0}_{0}=-\frac{2\sigma_0^2}{{N_0}^3}\delta N& 
-\frac{2}{{g_0^3}}\sigma'{_0}^{2}\delta g 
+\delta\sigma_1 \left(\frac{2}{{N_0}^2} {\sigma_0}^2 +\frac{2}{{g_0}^2} {{\sigma'}_0}^2 
+2(1+\frac{1}{2}\Lambda{\sigma_0}^2){\sigma_0}^2\right)
\\ \nonumber
&- \delta \dot \sigma_2\frac{2}{{N_0}^2}{\sigma_0}^2 + \frac{2}{{g_0}^2}\sigma_0{\sigma_0}'\delta{\sigma_1}',
\end{eqnarray}
\begin{equation}
\delta T^{1}_{1}=-\delta T^{0}_{0} + 4{\sigma_0}^2\delta\sigma_1,
\end{equation}
and
\begin{eqnarray}
\delta T^{2}_{2}=-\frac{2}{{N_0}^3}\sigma_0^2\delta N& 
-\frac{2}{{g_0^3}}\sigma'{_0}^{2}\delta g 
-\delta\sigma_1 \left(\frac{2}{{N_0}^2} {\sigma_0}^2 -\frac{2}{{g_0}^2} {{\sigma'}_0}^2 
-2(1+\frac{1}{2}\Lambda{\sigma_0}^2){\sigma_0}^2\right)
\\ \nonumber
&+ \delta \dot \sigma_2\frac{2}{{N_0}^2}{\sigma_0}^2 + \frac{2}{{g_0}^2}\sigma_0{\sigma_0}'\delta{\sigma_1}',
\end{eqnarray}
respectively.
Adding $\delta T^{0}_{0}$ to $\delta T^{1}_{1}$ we get
\begin{equation}
\left(\frac{{\delta N}'}{N_0}-\frac{{N_0}'\delta N}{{N_0}^2} - 
\frac{{\delta g}'}{g_0}+\frac{g'\delta g}{{g_0}^2} \right)= 
\frac{2}{{g_0}}\left(\frac{1}{r}+\left(\frac{{N_0}'}{N_0} - \frac{{g_0}'}{g_0}\right)\right)\delta g -
4 r {\sigma_0}^2 \left(1+\Lambda {\sigma_0}^2\right)\delta \sigma_1.
\end{equation}

Substituting equation (A20) as well as
for $\delta \dot \sigma_2$ from (A16) and (A19) in (A11)
we get
\begin{eqnarray}
\delta {\sigma_1}'' &+& \delta {\sigma_1}'\left(\frac{2}{r}
+\frac{{N_0}'}{N_0}-\frac{{g_0}'}{g_0}\right) + 
\frac{1}{{g_0}^2 r {\sigma_0}^2}\left(g_0\delta g' -{g_0}' \delta g\right)
-\frac{{g_0}^2}{{N_0}^2} \delta \ddot \sigma_1 
+ \frac{2\delta g}{g_0}\left[{\frac{{\sigma_0}'}{\sigma_0}}^2 
+ \frac{{g_0}^2}{{N_0}^2} \right.
\\ \nonumber
&+&\left.    \frac{1-2r\frac{{g_0}'}{g_0}}{2 r^2{\sigma_0}^2} -
{g_0}^2\left(1+ \frac{\Lambda {\sigma_0}^2}{2}\right)
+ \frac{{\sigma_0}'}{\sigma_0} \left(\frac{1}{r}+\frac{{N_0}'}{N_0}-\frac{{g_0}'}{g_0}\right)\right]
\\
\nonumber
&-& {g_0}^2\delta \sigma_1 \left[
\frac{1}{{N_0}^2}+\frac{1}{{g_0}^2}
 \left(\frac{{\sigma_0}'}{\sigma_0}\right)^2 +
\left(1+ \Lambda \sigma_{0}^{2}\right) 
+ 2 r \sigma_{0}'\sigma_0 \left(1 +\frac{\Lambda \sigma_{0}^2}{2}
\right)\right]=0.
\end{eqnarray}
Adding $\delta T_{0}^{0}$ to $\delta T_{2}^{2}$ and substituting in equation (A20) and
its derivative we get
\begin{eqnarray}
\frac{2}{g_0}{\delta g}'' &-& \frac{2g_0}{{N_0}^2}\ddot{\delta g} 
+ 8 \left(
2\sigma_0 {\sigma_0}' -r \left(1+\frac{1}{2}\Lambda {\sigma_0}^2\right){\sigma_0}^2{g_0}^2\right)
\delta {\sigma_1}'\\
\nonumber
&+& 8 \left[
2 {{\sigma_0}'}^2 -r 
{\sigma_0}^2{g_0}^2
\left(1+\frac{1}{2}\Lambda {\sigma_0}^2\right)
\left(\frac{2{\sigma_0}'}{\sigma_0}+\left\{\frac{2{N_0}'}{N_0}+\frac{{g_0}'}{g_0}\right\}\right)
\right]\delta \sigma_1\\ \nonumber
&-&\left[\frac{4{g_0}'}{{g_0}^2}+\frac{6}{g_0}\left(\frac{{N_0}'}{N_0}-\frac{{g_0}'}{g_0}\right)\right]
\delta g' + \frac{2\delta g}{g_0}\left[\frac{{g_0}''}{g_0}+2(\frac{{g_0}'}{g_0})^2
-3\frac{{g_0}'}{g_0}\left(\frac{{N_0}'}{N_0}-\frac{{g_0}'}{g_0}\right)
\right.\\ \nonumber
&-&\left.8 {{\sigma_0}'}^2-\frac{2}{r^2} - \frac{2}{{g_0}^2}\left(g_0{g_0}''-{{g_0}'}^2\right)
+2\left(\frac{{N_0}'}{N_0}-\frac{{g_0}'}{g_0}\right)^2+\frac{2}{r}
\left(\frac{2{N_0}'}{N_0}+\frac{{g_0}'}{g_0}\right)
\right]=0.
\end{eqnarray}
Using the expression for the particle number $N_p=\int_0^{\infty} d^3x J^{0}\sqrt{g}$ where
$J^{0}=\iota g^{00}\left(\phi_{,0}\phi^{\ast} -c.c. \right)$ we get
\begin{eqnarray}
\delta N_p =+4\pi\int^{\infty}_{0}\, &dr& \, {r}^2 \frac{g_0}{N_0}{\sigma_0}^2\, {\bf X} \left\{ 
 \frac{1}{{g_0}^2 r {\sigma_0}^2}\left(g_0\delta g' -{g_0}' \delta g\right)
\right. 
\\ \nonumber
&+& 
\frac{2\delta g}{g_0}\left({\frac{{\sigma_0}'}{\sigma_0}}^2 + \frac{1-2 r\frac{{g_0}'}{g_0}}{2
{r}^2{\sigma_0}^2} 
+ \frac{{g_0}^2}{{N_0}^2}\right) +
\frac{{\sigma_0}'}{\sigma_0}\delta \sigma '
\\ \nonumber
&-& \left. 
{g_0}^2\delta  \sigma_1 \left(\frac{1}{{N_0}^2}
 + \frac{1}{{g_0}^2}\left(\frac{{\sigma_0}'}{\sigma_0}\right)^2 +
\left(1+\frac{\Lambda}{2}
 \sigma_{0}^{2}
\right) \right)
\right\}.
\end{eqnarray}

In terms of bar coordinates defined in the beginning of this section equation (A21) becomes
\begin{eqnarray}
\frac{1}{\Lambda ^{1.5}}\delta {\bar \sigma_1}'' &+& \frac{1}{\Lambda^{1.5}}\delta {\bar \sigma_1}'\left(\frac{2}{\bar r}
+\frac{{N_0}'}{N_0}-\frac{{g_0}'}{g_0}\right) + 
\frac{1}{{g_0}^2 {\bar r} {\bar \sigma_0}^2}\left(g_0\delta g' -{g_0}' \delta g\right)
-\frac{1}{\Lambda^{0.5}}\frac{{g_0}^2}{{N_0}^2} \delta \ddot {\bar \sigma_1}
\\ \nonumber
&+& 
\frac{2\delta g}{g_0}\left(\frac{1}{\Lambda}
\left(
{\frac{{\bar \sigma_0}'}{\bar \sigma_0}}\right)^2 +
 \frac{{g_0}^2}{{N_0}^2} +
   \frac{1-2{\bar r}\frac{{g_0}'}{g_0}}{2 {\bar r^2}{\bar \sigma_0}^2} -
{g_0}^2\left(1+ \frac{ {\bar \sigma_0}^2}{2}\right)
+ \frac{1}{\Lambda}\frac{{\bar \sigma_0}'}{\bar \sigma_0} \left(\frac{1}{\bar r}+\frac{{N_0}'}{N_0}-\frac{{g_0}'}{g_0}\right)\right)
\\
\nonumber
&-&\frac{1}{\Lambda^{0.5}}{g_0}^2\delta \bar \sigma_1  \left(
\frac{1}{{N_0}^2}+
\frac{1}{\Lambda g_0^2}
\left ({\frac{{\bar \sigma_0}'}{\bar \sigma_0}}\right)^2 +
\left(1+  \bar \sigma_{0}^{2}\right) 
+ \frac{1}{\Lambda}2 {\bar r} \bar \sigma_{0}'\bar \sigma_0 \left(1 +\frac{ \bar \sigma_{0}^2}{2}
\right)\right)=0,
\end{eqnarray}
and equation (A22) becomes
\begin{eqnarray}
\frac{1}{\Lambda}\frac{2}{g_0}{\delta g}'' &-& \frac{2g_0}{N^2}\ddot{\delta g} 
+\frac{ 8}{\Lambda^{1.5}} \left(
\frac{2}{\Lambda}\bar \sigma_0 {\bar \sigma_0}' -r \left(1+\frac{1}{2} {\bar \sigma_0}^2\right){\bar \sigma_0}^2{g_0}^2\right)
\delta {\bar \sigma_1}'\\
\nonumber
&+& \frac{8}{\Lambda^{1.5}} \left[
\frac{2}{\Lambda} {\bar {{\sigma_0}'}^2} -r \left(1+\frac{1}{2} {\bar \sigma_0}^2\right){\bar \sigma_0}^2{g_0}^2
\left(\frac{2{\bar \sigma_0}'}{\bar \sigma_0}+\left(\frac{2{N_0}'}{N_0}+\frac{{g_0}'}{g_0}\right)\right)
\right]\delta \bar \sigma_1\\ \nonumber
&-&\frac{1}{\Lambda}\left[\frac{4{g_0}'}{{g_0}^2}+\frac{6}{g_0}\left(\frac{{N_0}'}{N_0}-\frac{{g_0}'}{g_0}\right)\right]
\delta g' + \frac{2}{\Lambda g_0}\left[\frac{{g_0}''}{g_0}+2(\frac{{g_0}'}{g_0})^2
-3\frac{{g_0}'}{g_0}\left(\frac{{N_0}'}{N_0}-\frac{{g_0}'}{g_0}\right)
\right.\\ \nonumber
&-&\left.\frac{8}{\Lambda} {\bar {{\sigma_0}'}^2}-\frac{2}{r^2} - \frac{2}{{g_0}^2}\left(g_0{g_0}''-{{g_0}'}^2\right)
+2\left(\frac{{N_0}'}{N_0}-\frac{{g_0}'}{g_0}\right)^2+\frac{2}{r}
\left(\frac{2{N_0}'}{N_0}+\frac{{g_0}'}{g_0}\right)
\right]\delta g=0.
\end{eqnarray}

Similarly equation (A23) becomes
\begin{eqnarray}
\delta N_p =4\pi\int^{\infty}_{0} &+& dr\, r^2\, \frac{g_0}{N_0}\,{\bar \sigma_0}^2\, {\bf X} \left\{ 
 \frac{1}{{g_0}^2 \bar r {\bar \sigma_0}^2}\left(g_0\delta g' -{g_0}' \delta g\right)
 \right.
\\ \nonumber
&+&
\frac{2\delta g}{g_0}\left(\frac{1}{\Lambda}({\frac{{\bar \sigma_0}'}{\bar \sigma_0}})^2 
+ \frac{1-2\bar r\frac{{g_0}'}{g_0}}{2
{\bar r}^2{\bar \sigma_0}^2} 
\frac{{g_0}^2}{{N_0}^2}\right) +
\frac{1}{\Lambda^{1.5}}\frac{{\bar \sigma_0}'}{\bar \sigma_0}\delta \bar \sigma '
\\ \nonumber
&-& \left. 
\frac{{g_0}^2}{\Lambda^0.5}\delta  \bar \sigma_1 \left(\frac{1}{{N_0}^2} 
+\frac{1}{{g_0}^2\Lambda} (\frac{{\bar \sigma_0}'}{\bar \sigma_0})^2 +
\left(1+ \frac{1}{2}
 \bar \sigma
_{0}^{2}\right) \right)
\right\}.
\end{eqnarray}
Using $\delta N_p =0$ (charge conservation) which is appropriate
for large $\Lambda$ where $\delta N_p$ is given by (A26) we get on putting this in (A24)
\begin{eqnarray}
\frac{1}{\Lambda ^{1.5}}\delta {\bar \sigma_1}'' &+& \frac{1}{\Lambda^{1.5}}\delta {\bar \sigma_1}'\left(\frac{2}{\bar r}
+\frac{{N_0}'}{N_0}-\frac{{g_0}'}{g_0}\right) + 
-\frac{1}{\Lambda^{0.5}}\frac{{g_0}^2}{{N_0}^2} \delta \ddot {\bar \sigma_1}
\\ \nonumber
&+& 
\frac{2\delta g}{g_0}\left(
-{g_0}^2\left(1+ \frac{ {\bar \sigma_0}^2}{2}\right)
+ \frac{1}{\Lambda}\frac{{\bar \sigma_0}'}{\bar \sigma_0} \left(\frac{1}{\bar r}+\frac{{N_0}'}{N_0}-\frac{{g_0}'}{g_0}\right)\right)
\\
\nonumber
&-&\frac{1}{\Lambda^{0.5}}{g_0}^2\delta \bar \sigma_1  \left(
\frac{ \bar \sigma_{0}^{2}}{2} 
+ \frac{1}{\Lambda}2 {\bar r} \bar \sigma_{0}'\bar \sigma_0 \left(1 +\frac{ \bar \sigma_{0}^2}{2}
\right)\right) -\frac{\bar \sigma_0'}{\Lambda^{1.5}\bar \sigma_0}\delta \bar \sigma=0.
\end{eqnarray}
This equation suggests that $\delta g$ goes like $\frac{1}{\Lambda^{0.5}}$. Writing $\ddot \delta g$
as $-\chi^2 \delta g$ and $\ddot \delta {\bar \sigma}$ 
as $-\chi^2 \delta \sigma$ we then see from(A25) and (A27) that the quasinormal mode frequency $\chi$ for
high $\Lambda$ configurations must go like $\frac{1}{\Lambda ^ {0.5}}$ for a given $\bar \sigma$. 
We have numerically evolved stars with the same $\bar \sigma$ and compared 
the QNM frequencies obtained
to the inverse ratios of the square root of their $\Lambda$ values for $\Lambda=800$, $\Lambda=1200$
and $\Lambda=1600$ confirming the above analysis.
In Table I we show the comparison between the perturbation analysis and the
numerical result. We see that the perturbation result gets more accurate for
increasing $\Lambda$.
We note that 
configurations that have the
size of a neutron star would have to have
$\Lambda$ of order $10^{38}$ with QNM frequency of order
$\chi = 10^{-19}$.

\footnotesize
\begin{center}
\begin{tabular}{|c|c|c|c|c|} 
\multicolumn{5}{c}{Table 1.}   \\
\hline

$\Lambda$ &
$1/f$ & $f_{1600}/f$ & $ \sqrt{\frac{\Lambda}{1600}} $
&
 \% error  
\rule[-0.1in]{0.0in}{0.3in}\\
 \hline
1600&       1220&         1&                   1&       
\rule[-0.1in]{0.0in}{0.3in}\\
 \hline
1200&       1070&        0.877&                  0.866&                   1.25
\rule[-0.1in]{0.0in}{0.3in}\\
 \hline
800 &        880&        0.7213&                 0.707     &              1.9
\rule[-0.1in]{0.0in}{0.3in}\\
 \hline
600 &        770&        0.6311&                 0.612 &                  3
\rule[-0.1in]{0.0in}{0.3in}\\
 \hline
\end{tabular}

\end{center}

\normalsize


\begin{figure}
\caption{The mass profiles of ground state boson stars for different values
of self-coupling constant $\Lambda$ are shown. The increase in mass with $\Lambda$ is clear although 
the profiles are very similar.}
\end{figure}
\begin{figure}
\caption{The comparison of a strongly perturbed
ground state $\Lambda=10$, $S$-branch star
[$ M=0.781 m_{Pl}^2/m, \sigma (0)=0.1 $] to the unperturbed
configuration [$ M=0.722 m_{Pl}^2/m$] is shown.
The solid lines correspond to the unperturbed configuration 
and the dashed ones to the perturbed star. 
The perturbation shown corresponds to  
the addition of scalar-field $\sigma$ at $t=0$.}
\end{figure}
\begin{figure}
\caption{(a) The evolution of the radial metric $g_{rr}=g^2$ for
the configuration shown in Fig.\ 2. The initial perturbed configuration is
labeled $t=0$. The unperturbed configuration is also shown. The spatial 
distributions of the radial metric labeled $A,B,C,D$
correspond to times $t=192,306,391,505$ respectively. The radial
positions of the peaks of the
radial metric for these times are labeled $R_A,R_B,R_C,$ and $R_D$. The
values $R_B$ and $R_D$ are so close that they appear as one thick line in
the figure.
(b) The peak value of the perturbed radial metric is plotted over a
long time. The points labelled $A,B,C,D$ correspond to the same labels in
$(a)$. The oscillations decay in time. (c) The total mass of the
star is plotted as a function of time. The mass loss through scalar
radiation decreases in time as the 
oscillations start damping out.}
\end{figure}
\begin{figure}
\caption{(a)The oscillation frequencies of different ground state boson star 
configurations are plotted as functions of mass, for 
$\Lambda=0, 5, 10, 15, 30, 100$ and $200$. 
The curves 
are obtained by slightly perturbing (perturbed mass within
$0.1\%$ of the unperturbed mass) $S$-branch stars.
They reach a peak and then drop down at the approach of the maximum mass
allowed for a given $\Lambda$ signalling a transition from stability to
instability. The frequencies for a given mass for higher $\Lambda$ stars
are lower than those for lower $\Lambda$ as a result of their
increased size. However their overall maximum frequencies could get bigger
than for lower $\Lambda$ stars because of their increase in maximum mass.
As can be seen $\Lambda=5$ and $\Lambda=10$ stars have higher
maximum frequencies than do $\Lambda=0$ stars. As the stars get very much larger
the maximum comes down as shown in the figure. $\Lambda=30$ stars have lower maximum frequency than
$\Lambda=15$ stars which have lower maximal frequency of oscillation than do
$\Lambda=10$ stars. 
(b)The highly perturbed $\Lambda=10$ $S$-branch 
star of figure 3, has an oscillation frequency below the
$\Lambda=10$ solid line. Its movement
towards the solid line is shown through points
$P1$, $P2$ and $P3$ corresponding to times 0, 1200,
and 4800 respectively.}
\end{figure}
\begin{figure}
\caption{(a) The radial metric $g^2=g_{rr}$ of a perturbed $U$-branch ground
state star is shown at various times. The curves
$1,2,3,a,b,c,d$ correspond to times $t=10,20,30,340,440,540$ and $640$. 
The unperturbed
star has a central density $\sigma(0)=0.23$ and a self coupling parameter
$\Lambda=30$. The initial equilibrium metric
configuration has also been shown.
The overall field  density of this star has been lowered by about
$10\%$. The $t=0$ curve corresponds to the initial perturbed
radial metric.
In the asymptotic
region $g^2$ is not oscillating but monotonically decreasing due to the mass
loss. 
(b) The maximum value of the radial metric is plotted as a function of
time. The initial sharp drop in the radial metric signifies the expansion of
the star as it proceeds to the stable branch. There it oscillates about
the new stable configuration that it is going to settle to. This corresponds
to a star of mass $M=1.037 m_{Pl}^2/m$ (whose
metric configuration has been shown in figure $5a$ as
a dark line). 
The points $a$ and $c$ correspond to two minima in the peak
of $g_{rr}$ which occur when the core of the star reaches its local maximum
size. Likewise the maxima in the peak of $g_{rr}$ at $b$ and $d$ correspond
to the core of the star reaching its local minimum size. (c) The mass is plotted against time.
The mass loss through scalar radiation at each expansion of the core 
(corresponding to the maximum radial metric reaching a minimum) decreases 
in time as the oscillations damp out.}
\end{figure}
\begin{figure}
\caption{The migration of the $U$-branch star considered in the previous figure
is shown after the star has moved to the $S$-branch. Points $Q1$, $Q2$, $Q3$ and $Q4$ correspond to
times 500,1000, 2000
and 3500 respectively. Here too the mass loss decreases in time and the star finally settles to
a stable configuration.}
\end{figure}
\begin{figure}
\caption{The masses of 0-node, 1-node and 2-node boson stars 
without self-coupling are plotted as a function of central
density. The maximum mass of 1-node stars is $1.356 m_{Pl}^2/m$ while the maximum
mass for 2-node stars is expectedly greater at $2.095 m_{Pl}^2/m$. 
The profiles are deceptively similar to their ground state counterpart. 
Excited state stars are
inherently unstable irrespective of the branch they lie on, unlike ground
state stars that can be termed stable or unstable depending on whether they
lie on the branch to the left of the maximum mass or to the right
respectively.}
\end{figure}
\begin{figure}
\caption{
(a)The transition of a first excited state star to the ground state is
shown. Here the radial metric is plotted against radius for various times
starting from $t=0$ and then for
$t_a=250,t_b=500 ,t_c=1245,t_d=4000,t_e=5000,t_f=5505,t_g=6000$ and $t_h=6370$. The initial
unperturbed and perturbed configurations
are shown. (The perturbed configuration
has the lower second peak.)
The initial mass of the star was $M=.901m_{Pl}^2/m$ after
perturbation. The radial metric 
initially has two peaks indicative of a 1-node configuration.
As the star evolves and goes to the
ground state one peak disappears. 
This can be seen in the curves from $t_c$--$t_h$. Once in
the ground state it oscillates and 
finally settles into a stable ground state configuration.
(b) The
transition of a 3 node configuration of
mass $0.91 m_{Pl}^2/m$. This star loses enough
mass during the course of its evolution to move to the 
ground state. 
}
\begin{figure}
\caption{
(a) A comparison of the manner of black hole formation of four exctited state configurations.
The first frame is an $S$ branch star of central density $\sigma(0)=0.1$ that tries
to go to the ground state but fails to. As the central density increases 
decays to black holes occur on a shorter time scale. A plot of the collapse of
a $U$ branch of central density $\sigma(0)=0.3$ is shown in
the next frame. Stars get more dispersive as one moves farther along the
$U$ branch. A star of central density $0.4$ shown in
the third frame has a decay time close to
the previous one. Decay times then start to increase. A star of
central density $0.5$ close to the $M=N\,m$ point ($\sim \sigma(0)=0.541$)
disperses to over ten times its radius before collapsing to a black hole.
(b) The decay time to black holes is plotted as a function of 
central density, for one node configurations (first excited states), 
of boson stars without self coupling. 
Perturbations are only due to the finite differencing effects of
the numerical scheme. To make the comparisons meaningful  the
$95\%$ mass radius (which is our definition for radius of a star)
of every
configuration considered was covered by the same number of grid points.
Configurations for which the central
density $\sigma(0) < \sigma_2$ move to the ground state. This value of the
critical density $\sigma_2=0.048$
corresponds to a mass $M=0.91m_{Pl}^2/m$. $\sigma_1=0.021$ is the value of
the central density corresponding to a mass $M=0.633m_{Pl}^2/m$, which is the
maximum mass of a ground state boson star.  The decay time 
decreases with increasing $\sigma(0)$ and this continues even for 
$\sigma(0) > 0.25$ which is the point of transition from the 
$S$-branch to the $U$-branch. 
The decay time then starts to increase as
one approaches the $M=N\,m$ point corresponding to a central density
$\sigma(0)=0.541$, beyond which 
the stars
disperse to infinity rather than become black holes. 
The dispersion times of two such stars to
ten times their original radius ($95\%$ mass radius) 
are also shown on the figure.
}
\end{figure}
\begin{figure}
\caption{(a) The evolution of the metric function $N^2=-g_{tt}$ for a
$\Lambda=30, \sigma(0)=0.1$ boson star in the first excited state, without
any explicit perturbation put, is shown. The configuration lies on the
$S$-branch and has an initial mass $M=1.743 m_{Pl}^2/m$. The various time slices
correspond to times $t=0,t_a=1060,t_b=1950$ and $t_c=1985$. The lapse function collapses
as an apparent horizon is approached, signaling the formation of the black
hole (indicative of an inherent instability of excited states).
(b) The evolution of the metric function $N^2=-g_{tt}$ for a
$\Lambda=0, \sigma(0)=0.122$ boson star in the first excited state, without
any explicit perturbation put, is shown. The configuration lies on the
$S$-branch and has an initial mass $M=1.23 m_{Pl}^2/m$. The various time slices
correspond to times $t=0,t_a=450,t_b=720,t_c=750$ and $t_d=770$. This star has a radius of
$20.7$ which is about the same as that of the configuration in (a). Again,
the lapse function collapses when an apparent horizon is approached, as a
black hole is being formed. The time scale of collapse is much less than for
the $\Lambda=30$ case in part (a).}
\end{figure}
\begin{figure}
\caption{
(a) The initial field configuration of a 5-node star.
The field has 5 nodes or extrema. The absolute value of each extremum is
clearly smaller than the one preceding it. 
(b) A contour plot of a perturbed 5-node star, that ends in a
black hole showing $\rho \times r^2$ 
as a function of distance (vertical axis) and of time (horizontal axis) is
shown. The density $\rho$ is highest at the origin and has five other local
maxima, each smaller than the previous one. The value of the maxima at the
end are very small compared to the earlier ones, and to enhance the features
$\rho \times r^2$ rather than just $\rho$ is shown in the plots. Each set
of lines represents the maxima of $\rho \times r^2$ and the number of
lines in a set gives an indication of the height of the maximum. 
This
particular configuration has a central density of 
$\sigma(0)=0.075$ and initial
mass $M=3.07m_{Pl}^2/m$.
This star cascades through an intermediate 4 node state (around $t=1000$)
before proceeding to form a black hole. Cascades are characteristic
of excited boson star decays similar
to atoms in excited states going through
intermediate states when transiting to the ground
state. 
(c) The equilibrium density function of a
4 node star of central density $\sigma=0.06$ (right frame) is 
placed alongside the
contour plot of the 5 node star described in $11b$ up to a time of $t=1000$
when it has
gone into a 4 node state
(left frame)
. This plot shows that the transition of the
5 node star is to a perturbed 4 node state close to the one shown in $11c$
before it continues its evolution to a black hole.}
\end{figure}
\end{figure}
\begin{figure}
\caption{ The equilibrium profiles of a $\Lambda=800$ star with central density $\sigma=0.05$ derived
from the high $\Lambda$ approximate equations and the exact ones are compared. A Schwarzschild
exterior is attached to the approximate solution after the field vanishes. The three plots show the
field $\sigma$, $g_{rr}$ and $g_{00}$ respectively. Clearly, the approximation matches the exact solution very
well.}
\end{figure}
\begin{figure}
\caption{ The mass of a high $\Lambda$ star generated from the approximate 
equations is plotted as a function of $\bar{\sigma}$ 
($\sigma/\Lambda^{\frac{1}{2}}$). It
shows the same basic structure as the profiles generated for low $\Lambda$
using exact equations. The peak is at about $.22 \Lambda  ^{1/2}m_{Pl}^{2}/m $
which means
to achieve $ 0.1 \msun $ would take $\Lambda$ 
of order $10^{38}$ a very large star
to evolve numerically. Also plotted is $N\, m$ ($N$ is the particle number
and $m$ the mass of a boson). The crossing point 
of the two curves represents
transition from negative to positive binding energy.}
\end{figure}

\begin{table}
\vspace{0.4in}
\caption{The ratio of the QNM frequency for $\Lambda=1600$ 
to the QNM frequency for a given $\Lambda$
is compared to $\frac{ \Lambda ^{0.5}}{40}$ 
(which is the predicted ratio for large $\Lambda$)
for $\Lambda = 1200$, $800$ and $1600$. The higher $\Lambda$ values match better as
expected.
The initial central density is $\bar \sigma (0) =0.4.$
}
\label{tbh}
\end{table}

\begin{thebibliography}{10}

\bibitem[1]{pri} J.R. Primack, D. Seckel and B. Sadoulet,
Annu.\ Rev.\ Nucl.\ Part.\ Sci.{\bf 38}, 751 (1988).

\bibitem[2]{sei}E. Seidel and W-M. Suen, 
Phys.\ Rev.\ Lett.{\bf 66}, 2516 (1993).


\bibitem[3]{lee} T. D. Lee and Y. Pang,
Phys.\ Rep.{\bf 221}, 251 (1992) 
and 
A. R. Liddle and M. S. Madsen , 
Int. J. Mod. Phys. D, {\bf 1}, 101 (1992).


\bibitem[4]{jet}
P.Jetzer, Phys.\ Rep.{\bf 220}, 183 (1992).

\bibitem[5]{sei}E. Seidel and W-M. Suen, 
Phys.\ Rev.\ D {\bf 42}, 384 (1990).

\bibitem[6]{stu} M. Colpi, S.L. Shapiro and I. Wasserman,
Phys.\ Rev.\ Lett.{\bf 57}, 2485 (1986).

\bibitem[7]{bre} J.D. Breit, S. Gupta and A. Zaks,
Phys.\ Lett.{\bf 140B}, 329 (1984).

\bibitem[8]{gle} M. Gleiser,
Phys.\ Rev.\ D {\bf 38}, 2376 (1988).

\bibitem[9]{jet} P. Jetzer,
Fermilab Report No. Conf-88/88-A (unpublished)

\bibitem[10]{lee} T. D. Lee,
Phys.\ Rev.\ D {\bf 35}, 3637 (1987).

\bibitem[11]{gle} M. Gleiser and R. Watkins,
Nucl.\ Phys. {\bf B319}, 733 (1989).

\bibitem[12]{lee} T. D. Lee and Y. Pang,
Nucl.\ Phys. {\bf B315}, 477 (1989).

\bibitem[13]{ruf} R. Ruffini and S. Bonozzola,
Phys.\ Rev.{\bf 187}, 1767 (1969).

\bibitem[14]{alc}Alcock et al , 
Int. Astron. Union Circ.(USA)
{\bf 6068}, (1993).

\bibitem[15]{bar}J.M. Bardeen and T. Piran, 
Phys.\ Rep.\  {\bf 96}, 205 (1983).


\bibitem[16]{Bon}C. Bona, 
Private Communications. 

\end{thebibliography}
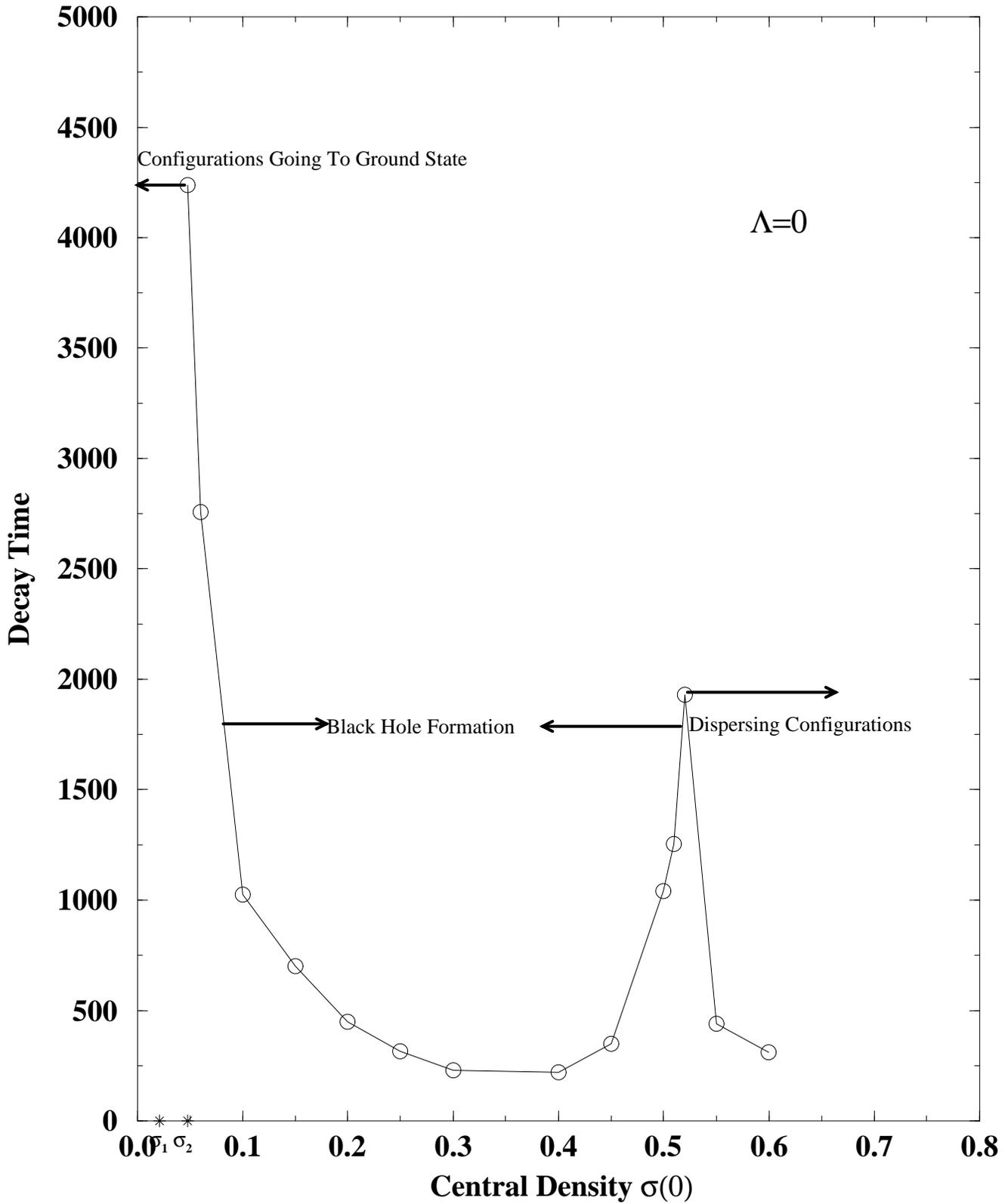
\end{document}